\documentclass[aps,prl,twocolumn,superscriptaddress]{revtex4-1}
\usepackage{graphics}
\usepackage{graphicx}
\usepackage{color}
\usepackage{epstopdf}
\usepackage{amsmath}
\usepackage{cancel}
\begin{document}
\title{Controlling Marangoni induced instabilities in spin-cast polymer films: \\ how to prepare uniform films} 
\author{Paul D. Fowler}
\affiliation{Department of Physics and Astronomy, McMaster University, 1280 Main St. W, Hamilton, ON, L8S 4M1, Canada.}
\author{C\'{e}line Ruscher} 
\affiliation{Department of Physics and Astronomy, McMaster University, 1280 Main St. W, Hamilton, ON, L8S 4M1, Canada.}
\affiliation{Institut Charles Sadron, Universit\'{e} de Strasbourg, Strasbourg, France}
\author{Joshua D. McGraw}
\affiliation{D\'epartement de Physique, Ecole Normale Sup\'erieure / PSL Research 
University, CNRS, 24 rue Lhomond, 75005 Paris, France}
\author{James A. Forrest}
\affiliation{Department of Physics \& Astronomy, University of Waterloo,  Waterloo, Ontario, Canada, N2L 3G1} 
\author{Kari Dalnoki-Veress}
\email{email: dalnoki@mcmaster.ca}
\affiliation{Department of Physics and Astronomy, McMaster University, 1280 Main St. W, Hamilton, ON, L8S 4M1, Canada.}
\affiliation{Laboratoire de Physico-Chimie Th\'eorique, UMR CNRS Gulliver 7083, ESPCI Paris, PSL Research University, 75005 Paris, France.}

\begin{abstract}
In both research and industrial settings spin coating is extensively used to prepare highly uniform thin polymer films.  However, under certain conditions, spin coating results in films with non-uniform surface morphologies. Although the spin coating process has been extensively studied, the origin of these morphologies is not fully understood and the formation of non-uniform spincast films remains a practical problem. Here we report on experiments demonstrating that the formation of surface instabilities during spin coating is dependent on temperature. Our results suggest that non-uniform spincast films form as a result of the Marangoni effect, which describes flow due to surface tension gradients. We find that both the wavelength and amplitude of the pattern increase with temperature. Finally, and most important from a practical viewpoint, the non-uniformities in the film thickness can be entirely avoided simply by lowering the spin coating temperature.
\end{abstract}

\maketitle
\section{Introduction}
\label{intro}
\begin{figure}[t]
\begin{center}
\includegraphics[width = 3.5in]{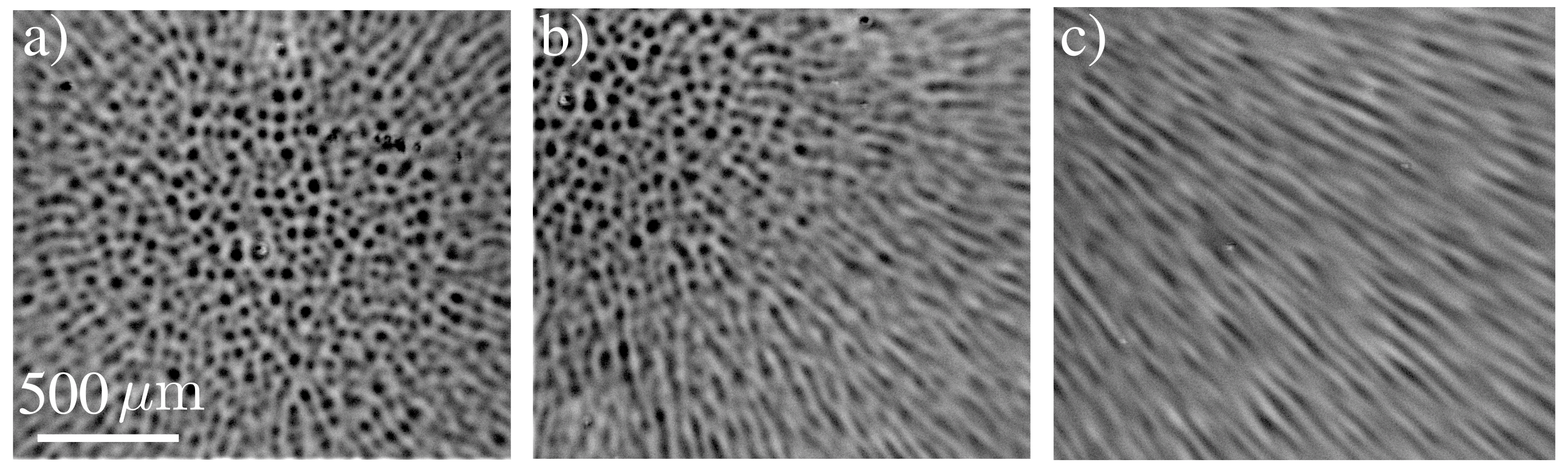}
\caption{Optical microscopy images of the typical `flowering morphology' observed in our experiments. In (a), the centre of the image corresponds to the axis of rotation of the spincoater, which we refer to as the `spinning centre'. Near the spinning centre, the pattern is isotropic and has a well defined wavelength. (b), (c) Farther from the spinning centre, the morphology turns to striations in the film, which point radially outward from the spinning centre. In the images the variation in intensity results from the interference of light, thus reporting on variations in height. Images are taken from a polystyrene film ($ M_w =  183 \; \text{kg/mol}$) spin-cast from a toluene solution at $T = 35^{\circ}\text{C}$.}
\label{flower}
\end{center}
\end{figure}

Spincoating is a widely used technique for the production of uniform thin films and has a diverse range of industrial applications including biomedical coatings, microelectronics, and solar cell technology \cite{Norrman:2005dk,Sirringhaus:1998gc,Walheim:1999ed}. The technique involves a small amount of polymer dissolved in a volatile solvent. This solution is deposited onto a flat substrate which is then rapidly spun (or which was already spinning) causing most of the solution to be ejected from the substrate. The remaining solvent evaporates leaving behind a polymer film with thicknesses typically on the order of nanometres to microns.  In many circumstances spincoating produces highly uniform films, however under certain conditions the process results in films with non-uniform surface morphologies, such as the one shown in fig. \ref{flower}. Because of its prevalence as a preparation technique, the spincoating process has been widely studied and much effort  has gone into characterizing the factors  that give rise to such morphologies \cite{deGennes:2002jl,deGennes:2001wz,Birnie:2011js,MullerBuschbaum:2001dz,MullerBuschbaum:2000ef,MullerBuschbaum:1998uy,Strawhecker:2001ch,Wang:1995tl,BirnieIII:2013cu}. Although much is known about the  spincoating of films, the formation of non-uniform spincast films remains a practical problem in both industrial settings and research laboratories.  

One of the first successful theoretical models of spincoating was developed by Emslie \emph{et al.} \cite{Emslie:1958et} who solved the equations of motion for a non-volatile Newtonian fluid on a rotating disk, by equating centrifugal and viscous forces. Although this model captures the fundamental physics of spincoating, the process is usually far more complicated. In practice, films are typically spincast from polymer solutions, meaning surface tension effects, shear thinning and elasticity must be considered, for example. Furthermore, as spinning proceeds, the properties of the solution change due to solvent evaporation. 

Later models incorporated some of these effects. In \cite{Flack:1984is}, Flack \emph{et al.}, proposed a model which includes the non-Newtonian properties of the solution as well as changes in the solution viscosity due to solvent evaporation. This model breaks the spincoating process into two stages, the first controlled by viscous radial flow of the solution and the second dominated by solvent evaporation. One major success of this model is its prediction of the experimentally observed dependence of film thickness on spin speed. Later work \cite{Lawrence:1988ci,Lawrence:1990dv,Lawrence:1991vl,Bornside:1989hq} further investigated the effects of solvent evaporation on the properties of spin-cast films. In \cite{Lawrence:1990dv,Lawrence:1991vl} the authors suggest that the concentration of solvent in the film is constant except in a  boundary region adjacent to the solution/air interface which is thin compared to the thickness of the solution. For the cases of both slow and rapid solvent evaporation, it is found that the final thickness of the spin-cast film depends on the thickness of the boundary layer. 

Bornside \emph{et al.} proposed a similar model \cite{Bornside:1989hq} which incorporates variations in viscosity and diffusivity across the thickness of the film. It was found that under certain conditions, a region of low solvent concentration develops at the free surface. Since this surface layer has a high viscosity and low diffusivity, it behaves like a `solid skin' which retards solvent evaporation. The authors propose that if solvent evaporation is rapid enough, the solid skin will develop while there is still flow in the liquid below, leading to hydrodynamic instabilities. This, in turn, causes inhomogeneities in the film. Such morphologies have been observed in the literature, \cite{Birnie:2011js,MullerBuschbaum:2001dz,MullerBuschbaum:2000ef,MullerBuschbaum:1998uy,Strawhecker:2001ch,Spangler:1990ux,Bormashenko:2006,Bormashenko:2010} leading to many further studies on the formation of instabilities during spincoating \cite{Wang:1995tl,Reisfeld:1991gh,Craster:2009ca,Oron:1997vo,Munch:2011dx}. In particular, de Gennes proposed a model to describe the formation of a solid skin or `crust' which includes an estimate for the minimum film thickness required to form the crust \cite{deGennes:2002jl,deGennes:2001wz}. In \cite{deGennes:2002jl} de Gennes argues that upon drying the crust is under mechanical tension and can rupture leading to film thickness variations.  Some authors have argued against crust formation, instead suggesting the formation of non-uniform films is driven by the Marangoni effect \cite{Birnie:2011js,Strawhecker:2001ch,BirnieIII:2013cu,Bassou:2009dp}. To date, the problem remains unsolved and due to the complexity of the spincoating process, efforts to combat the formation of non-uniform films are largely empirically based.

Experimentally, the spincoating process has been extensively studied. It is well known that the properties of a spin-cast film depend on numerous factors including spin speed, solution concentration, solution viscosity and vapour pressure \cite{Norrman:2005dk,Spangler:1990ux,Meyerhofer:1978ho}. More recent literature highlighting the complexity of the process has shown that spincoating may also affect fundamental material  properties of the film such as the entanglement network and viscosity \cite{Barbero:2009js,Thomas:2011kv,Raegen:2010ck,Reiter:2005br,McGraw:2013el}. Despite significant advancements in the understanding of spincoating, it remains an outstanding goal to control the morphology of thin polymer films. One well studied case is that of phase separation in spin-cast polymer blends~\cite{jukes,mokarian}. In addition to characterizing how the morphology depends on factors such as polymer concentration and spin speed \cite{DalnokiVeress:1997vn,Walheim:1997vp,Heriot:2005ie,Toolan:2013en}, more complex effects have been investigated, including  confinement \cite{DalnokiVeress:1998un} and the patterning of substrates \cite{Boltau:1998vl}. Pattern formation in spin-cast diblock-copolymer films has also received significant attention \cite{Green:2001wt,Kim:2004gd}. Furthermore, many authors have studied the morphology of films spin-cast from complex solutions such as colloidal suspensions and sol-gels \cite{Birnie:2011js,Taylor:2002gl,Rehg:1992uo,Zhao:2008bh}.

Here, we are primarily focused on the simple and common case of spincoating a thin film from a poly\-sty\-rene/tol\-uene solution. As mentioned above, non-uniformities are typically attributed to either the formation of a crust or the Marangoni effect. In the case of a Marangoni process, gradients in surface tension drive the formation of convection cells, which ultimately lead to variations in the film thickness. Surface tension gradients may arise due to gradients in temperature or concentration. In either case, once the surface tension gradient has been established, solution begins to flow from the region of low surface tension, toward the area of high surface tension. As liquid flows away from the low surface tension regions, fluid must come in to take its place. Under certain conditions, this effect is amplified and convection cells form. The Marangoni effect has been used to explain pattern formation in spin-cast films and in films formed through the drying of a polymer solution \cite{Strawhecker:2001ch,BirnieIII:2013cu,Bormashenko:2006,Bormashenko:2010,Bassou:2009dp,Sakurai:2002wa}. 

The details of the Marangoni effect have been known for years \cite{Craster:2009ca,Oron:1997vo,Block:1956wv,Scriven:1960go,Pearson:1958wl,Schatz:2001tk,Johnson:1999bm} but to date there are few experimental techniques to combat the formation of non-uniform spin-cast films. In \cite{Bornside:1989hq} it was suggested that saturating the environment above the solution/air interface with solvent may help to prevent the development of instabilities. Although this technique is routinely used in research laboratories, the setup can be cumbersome and its success depends on the particular polymer/solvent combination. Other authors have proposed that by appropriately mixing solvents, instabilities can be avoided \cite{MullerBuschbaum:2001dz}. Though also practiced in laboratories, this approach can be tedious and depends on many factors such as: the molecular weight of the polymer, solvent solubilities and ratios, and the tendency for phase separation during spin coating of the multiphase solvent.  

In this work we have systematically studied the effects of temperature on the formation of non-uniformities in spin-cast polymer films. We find that the dominant wavelength and amplitude of the morphology is dependent on the spinocating temperature.  Our results are consistent with a Marangoni process driving the non-uniform film formation, as discussed below.  From a practical perspective, we find that \emph{non-uniformities in the film thickness can be avoided simply by decreasing the spin-coating temperature}.

\section{Experiment}
\label{expt}

Samples were prepared using polystyrene (PS) with nine different weight averaged molecular weights, $8.8 \leq M_w \leq 758.9 \; \text{kg/mol}$, each with a polydispersity index less than 1.10 (Polymer Source, Inc.). Experiments were also performed using a symmetric diblock co-polymer, poly\-(styrene-methyl methacrylate) (PS-PMMA) with a total molecular weight of 211 kg/mol and a polydispersity index of 1.13 (Polymer Source Inc.). Each polymer was dissolved into toluene (Fisher Scientific, Optima grade) in various weight fractions ranging from $\phi = 1.5$ to 4.5 wt\%; in the following we use, \emph{e.g.} 2.5\,\% to mean 2.5 percent by weight. Films were spincast onto clean $20\; \text{mm} \times 20 \; \text{mm}$ Si wafers (University Wafer) using a spin speed of $4000 \; \text{rpm}$. The resulting films ranged in thickness from 100 to 200~nm.  

Spincoating was performed using a simple setup which allowed control of the spincoating temperature. Si wafers were placed onto a large home made aluminum disk, which we refer to as the spincoater chuck, and held in place using small screws. With screws holding the substrate onto the chuck good thermal contact is ensured. The temperature of the chuck was adjusted by placing it on either a heater or in a cooled environment. A thermocouple embedded well within the chuck\footnote{The thermocouple is inserted through a small hole (diameter $\approx 1 \;\text{mm}$) that enters  deep into the side of the chuck. The contact of the thermocouple well within the large thermal mass ensures an accurate temperature reading as well as easy insertion and removal of the thermocouple.} was used to monitor the temperature, which ranged from $15 \leq T \leq 60 \, ^{\circ} \text{C}$.  Once the chuck reached the desired temperature, it was transferred onto a commercial spincoating apparatus (Headway Research Inc., Model PWM32). Several drops of the polymer/toluene solution were placed onto the wafer and the chuck was rotated at 4000 rpm. Immediately prior to spinning, the chuck temperature was recorded and the thermocouple was removed from the chuck.

\begin{figure}[t]
\begin{center}
\includegraphics[width = 3in]{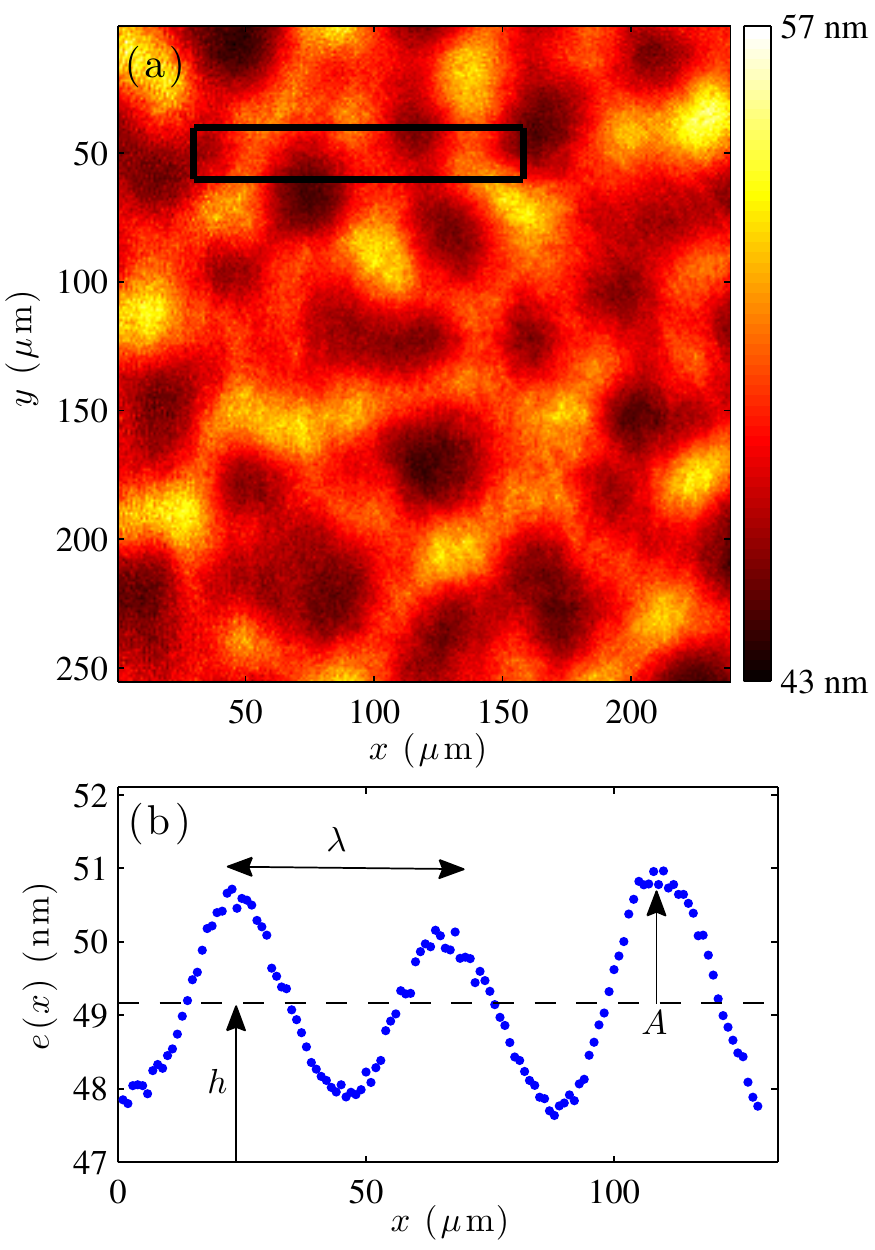}
\caption{(a) Imaging ellipsometry measurement of the topography near the spinning centre of a PS-PMMA film spin-cast from a $\phi = 3.5\,\%$ solution at room temperature, $T_{\mathrm RT} = 22^{\circ}\text{C}$. (b) Typical height profile of a film. The profile is an average over the region contained in the black box shown in (a). The amplitude ($A$), the average film thickness ($h = \langle e(x) \rangle$) as indicated by the dashed line, and wavelength ($\lambda$) are defined in the plot. }
\label{ep3}
\end{center}
\end{figure}

\begin{figure}[t]
\begin{center}
\includegraphics{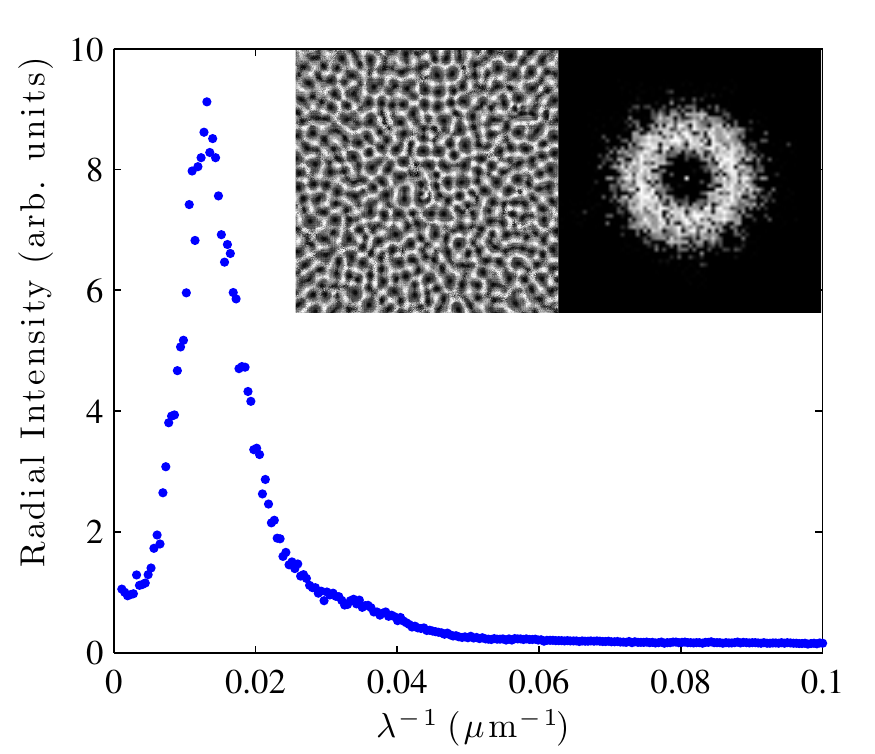}
\caption{Radial intensity as a function of inverse wavelength obtained from the FFT and contrast enhanced OM image shown in the inset. The bright ring in the FFT, which corresponds to the characteristic wavelength in the image appears as a sharp peak in the radial intensity. The data are taken from a PS(183k) film spin-cast from a solution with $\phi = 2.5\,\%$ at $T = 35^{\circ}\text{C}$. }
\label{fft}
\end{center}
\end{figure} 

Since the substrate is in good thermal contact with the chuck and the thermal mass of the solution is negligible compared to the chuck/substrate, we assume that the temperature of the solution is that of the chuck at the start of spinning. To test this, a second thermocouple was used to probe the temperature of the solution immediately following its deposition onto the substrate. The temperature difference between the chuck and solution was less than $1^{\circ}$C, which is within experimental error. Having confirmed that the temperature of the solution can be assumed to be that of the chuck, for all subsequent experiments only the temperature of the chuck at the start of spinning was recorded. We emphasize that spincoating was carried out under typical spincoating conditions: at ambient temperature and in air. Throughout this work the spincoating temperature refers to the temperature of the spincoater chuck at the onset of spinning. 

In fig. \ref{flower} are shown optical microscopy (OM) images of the typical `flowering morphology' observed in our experiments. The polymer-solvent combination of PS and toluene typically results in uniform films at ambient conditions. However, at the elevated spincoating chuck temperature of 35$^\circ \mathrm{C}$ the `flowering morphology' is observed. The variations in the intensity of the greyscale OM image result from the varying reflectivity of the thin transparent film atop the reflective substrate. Thus, the variations in the intensity are related to varying film thickness. Near the centre of the film, the pattern is isotropic with a well defined wavelength. Toward the edge of the film, this pattern turns into radial variations in the film thickness. To characterize this morphology, samples were imaged using both OM and imaging ellipsometry (IE, Accurion, EP3). In this work we focus on the isotropic region near the centre of film and investigate how the amplitude and wavelength of this pattern change with temperature.  

\emph{Amplitude:} Fig. \ref{ep3}(a) shows a typical map of the topography in the central region of the film as measured with imaging ellipsometry. From images such as these we are able extract a height profile of the film. In fig. \ref{ep3}(b) is shown an example of a height profile, where we have defined $h$ as average film thickness: $h = \langle e(x) \rangle$. The amplitude ($A$) is defined as half the height difference between the maximal and minimal film thicknesses. The wavelength ($\lambda$) is defined as the average distance over which the pattern repeats. For all data shown below, the reported amplitude is an average over multiple ellipsometry measurements of at least 5 samples. 

\emph{Wavelength:} The wavelength was calculated using a two-dimensional Fast Fourier Transform (FFT). The OM image of the central region of the film was contrast enhanced prior to the FFT in order to increase the signal-to-noise ratio. This process does not affect the characteristic length scale of the flowering pattern, which is the physical quantity of interest. The inset of fig. \ref{fft} shows a typical optical image, after contrast enhancement along with its FFT. The bright ring seen in the FFT corresponds to the characteristic frequency in the image. By plotting the radial average of the intensity as a function of inverse wavelength, we are able to determine the wavelength of the pattern.  In fig. \ref{fft} is shown the plot of radial intensity as a function of inverse wavelength which corresponds to the optical image and FFT shown the inset of the figure.  For all samples analyzed in our experiments, there is a sharp peak in the radial intensity of the FFT, resulting in a well defined wavelength. For all data shown below, the reported wavelength represents the average value of measurements taken from 5 to 10 films. 

\section{Results and discussion}
This section is divided into three parts. In sections~\ref{wavelength} and \ref{amplitude} we present the results of experiments performed with  spincoating temperatures above ambient temperature. We have systematically studied how both the wavelength and amplitude of the flowering morphology change with increasing spincoating temperature. We establish that the formation of the flowering pattern is consistent with a Marangoni driven process. Finally, in section \ref{cooling} we discuss experiments performed with spincoating temperatures below ambient temperature. We show that by spincoating at cooler temperatures we are able to suppress the formation of non-uniform films. 
\subsection{Wavelength}
\label{wavelength}
\begin{figure}[t]
\begin{center}
\includegraphics{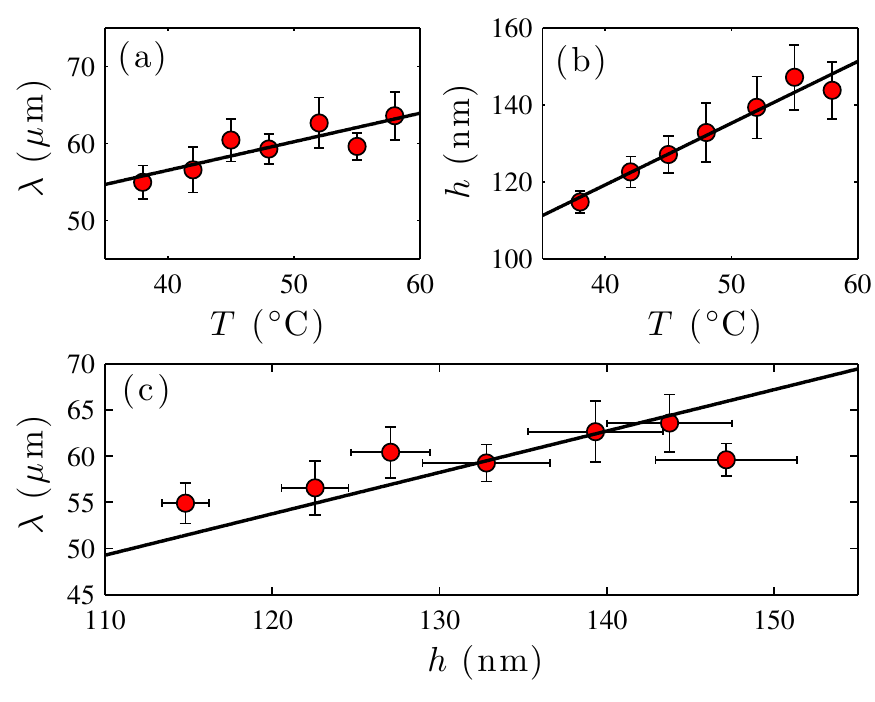}
\caption{ Measurements of the wavelength and average film thickness as a function of spincoating temperature for films spin-cast from a solution of PS(183k) with $\phi = 2.5$\,\%. Each data point represents an average of measurements taken from many films and the distance from the top to bottom of an error bar is one standard deviation.  (a) Plot of wavelength as a function of spincoating temperature with best fit line. (b) Plot of average film thickness as a function of spincoating temperature with best fit line. (c) Plot of wavelength as a function of average film thickness, obtained from the data in (a) and (b). The data is fit to a straight line forced through the origin as predicted by the Marangoni instability (see text). In Supplementary Material, fig. S1, we show that the slope $d\lambda/dh \approx 0.5$ is independent of the PS molecular weight for all the PS studied here with $\phi = 2.5$\,\%. }
\label{lambdas1}
\end{center}
\end{figure}

In fig. \ref{lambdas1} is shown the results of experiments for films spin-cast from a solution of PS(183k) with $\phi = 2.5\,\%$ as a function of the spincoating temperature, $T$. Fig. \ref{lambdas1}(a) shows a plot of the wavelength as a function of spincoating temperature. In fig. \ref{lambdas1}(b) is shown the average film thickness plotted as a function of the spincoating temperature. We find that both the wavelength of the flowering morphology and the average film thickness increase linearly with spincoating temperature. Combining the data in figs. \ref{lambdas1}(a) and (b), we plot the wavelength  as a function of the average film thickness in fig. \ref{lambdas1}(c). 

As mentioned in section \ref{intro}, non-uniformities in spin-cast polymer films are typically attributed to either the formation of a crust at the free surface, or the Marangoni effect \cite{deGennes:2002jl,deGennes:2001wz,Strawhecker:2001ch,BirnieIII:2013cu}. For a Marangoni process it is predicted that the wavelength scales with the average film thickness: $\lambda \propto h$~\cite{Oron:1997vo,Bassou:2009dp,Pearson:1958wl}. For the crust formation mechanism there is no prediction of an emergent wavelength~\cite{deGennes:2002jl,deGennes:2001wz}. Since the data shown in fig. \ref{lambdas1}(c) is well described by a straight line through the origin, we conclude that our wavelength data is consistent with the predicted scaling for a  Marangoni process~\cite{Oron:1997vo,Bassou:2009dp,Pearson:1958wl}.  

Marangoni convection is driven by surface tension and there are two clear mechanisms which can result in surface tension gradients across the polymer solution. First, since the solution is heated from below by the spincoater chuck with evaporative cooling at the free interface, there is a temperature gradient across the solution. If there is any perturbation to the free surface, the region of solution which is closer to the heated chuck will be hotter than the fluid which has been pushed away from the substrate.  Since surface tension decreases with temperature, the regions of interface which are closer to the heated chuck will have a lower surface tension compared to those farther from the chuck. Such a  gradient in surface tension can drive flow. High surface tension regions pull fluid along the interface away from the areas of low surface tension. As solution moves away from areas of low surface tension, fluid from below flows to take its place. Normally, this surface tension driven flow is mediated by viscosity and thermal diffusion and the liquid flattens. However, under certain conditions, the flow induced by surface tension gradients is amplified leading to the formation of convection cells in the liquid.  The balance between surface tension forces and dissipation due to thermal diffusivity and viscosity is characterized by the dimensionless Marangoni number,

\begin{equation}
	M = - \frac{d\gamma}{dT}\frac{h\Delta T}{\eta \alpha}\ ,
	\label{marangoni}
\end{equation}
where $h$ is the thickness of the fluid, $\alpha$ the thermal diffusivity and $\Delta T$ the temperature gradient across the fluid. It has been shown that in order for a Marangoni instability to occur, $M$ must exceed some critical value, $M_c$ \cite{Pearson:1958wl,Schatz:1995tk}.

The second mechanism by which Marangoni convection cells can form is as a result of concentration gradients induced by rapid solvent evaporation. Concentration gradients along the free surface establish surface tension gradients which drive the formation of convection cells via the mechanism described above. Due to the complexity of our system, we cannot determine whether the Marangoni process is triggered by gradients in temperature or concentration and it may be the combination of the two effects. 

To summarize, for a Marangoni process the wavelength is predicted to scale with the average film thickness~\cite{Oron:1997vo,Bassou:2009dp,Pearson:1958wl}. We find that the data shown in fig. \ref{lambdas1}(c) is consistent with this prediction as shown by the fit line (forced through the origin).  We note that the measurements shown in fig. \ref{lambdas1} (a) and (b) were repeated with varying solution concentrations ranging from $\phi=1.5\,\%$ to $\phi=4.5\,\% $ and for nine PS molecular weights ranging from 8.8~kg/mol to 759~kg/mol (see Electronic Supplementary Material, Table 1). In all cases, the same trend exemplified in fig. \ref{lambdas1} (a) and (b) was observed: the wavelength and average film thickness increase linearly with temperature.

\subsection{Amplitude}
\label{amplitude}
\begin{figure}[t]
\begin{center}
\includegraphics{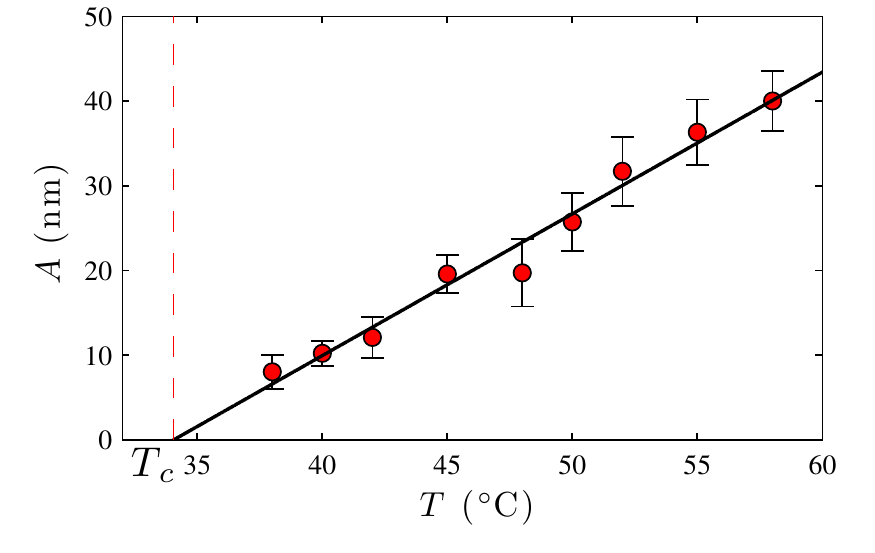}
\caption{ Measurements of the amplitude as a function of spincoating temperature for films spin-cast from a solution of PS(183k) with $\phi = 2.5$\,\%. Each data point represents an average of many measurements taken with IE and the distance from the top to bottom of an error bar is the standard deviation. The critical temperature $T_c$ is defined by the intersection of the best-fit straight line with the $T$ axis, as indicated by the vertical dashed line. As with $d\lambda/dh$, $T_c$ is also independent of molecular weight for a given concentration, as shown in fig. S2. By contrast, fig. S3 shows that $dA/dT$ is molecular weight dependent and $T_c$ is concentration dependent, fig. S4. }
\label{amp1}
\end{center}
\end{figure}

In the previous section, we established that the formation of the flowering morphology is consistent with a Marangoni process. We now present the results of experiments investigating the effect of temperature on the amplitude of the flowering pattern. In fig. \ref{amp1} is shown a plot of the amplitude as a function of spincoating temperature for films spincast from a $\phi = 2.5\,\%$ solution of PS(183k). The data in fig. \ref{amp1} shows that the amplitude increases linearly with the spincoating temperature. Furthermore, extrapolation of the data reveals that below $T\approx 34^\circ \mathrm{C}$, the flowering morphology vanishes -- an observation that is consistent with the fact that PS-toluene is a common system with which uniform films are prepared at ambient conditions. To explain this trend, we compare with the results of recent studies on the formation of non-uniformities in spin-cast polymer films. 
In work by  M\"uller-Buschbaum \emph{et al.} the morphology of films spin-cast from various solvents was studied~\cite{MullerBuschbaum:2001dz}. Qualitatively, the patterns observed in \cite{MullerBuschbaum:2001dz} appear similar to those presented in this work. The authors found that the surface roughness of the films increased with increasing vapour pressure. Strawhecker \emph{et al.} also studied morphologies similar to the flowering patterns seen here and found that the surface roughness increased as a function of vapour pressure \cite{Strawhecker:2001ch}. To explain this result, the authors propose that the morphologies form during the early stages of spincoating and depend on the competition between two phenomena: 1) The temperature gradient induced by rapid solvent evaporation leads to Marangoni instabilities which roughen the surface. 2) The solution is driven to level in order to minimize the surface energy. To describe the competition between these two effects, the authors define the ratio between the levelling time and evaporation time as \cite{Strawhecker:2001ch}

\begin{equation}
	\Lambda =\frac{\tau_{\mathrm{level}}}{\tau_{\mathrm{evap}}} \propto  \frac{\eta E}{\gamma \rho \theta^{m+1}}\ ,
	\label{kumar}
\end{equation}
where $\eta$, $E$, $\gamma$ and $\rho$ are the viscosity, evaporation rate, surface tension and density of the solvent. $\theta$ is the contact angle of a drop of solution and $m$ a positive exponent. During the early stages of spincoating, the dilute solution has a low viscosity and there is a competition between the levelling of the surface and the formation of Marangoni induced roughness. As solvent evaporates, the viscosity increases and eventually the film `freezes'. If the film vitrifies before the surface has time to level, the resulting film will be non-uniform. Increasing the evaporation rate, reduces the time the film has to flatten before it vitrifies, which leads to rougher films. The authors note that since evaporation rate increases with vapour pressure this simple picture is consistent with their experiment results. This model also explains the trend of increasing roughness with increasing vapour pressure observed in \cite{MullerBuschbaum:2001dz}.
Similarly, the results shown in fig. \ref{amp1}(a) are consistent with the ideas presented in \cite{Strawhecker:2001ch}. Increasing the spincoating temperature leads to more rapid solvent evaporation and therefore rougher films. We also note that increasing the spincoating temperature increases both surface tension gradients and the temperature gradient across the polymer solution while lowering the solution viscosity. Therefore raising the spincoating temperature increases the convection velocity and the driving force for Marangoni instabilities. Thus, according to eq. \ref{marangoni} the incerease in the amplitude is consistent with the increasing value of $M$ as the spincoating temperature is increased. 

Finally, as mentioned in section~\ref{wavelength} a second mechanism often used to explain the formation of spincoating instabilities is crust formation. According to de Gennes~\cite{deGennes:2002jl,deGennes:2001wz}, a solvent depleted region near the free surface results in a solid skin or `crust' which can rupture if it is sufficiently thin leading to film thickness variations. For the crust formation mechanism we expect the amplitude of the thickness variations to increase with crust thickness and there is no prediction for an emergent wavelength. In~\cite{deGennes:2002jl} de Gennes shows that with increasing vapour pressure, the crust thickness decreases. We note that since the amplitude of the flowering pattern increases with temperature and thus vapour pressure our results are not consistent with the rupture of a crust leading to thickness variations, which provides further evidence for the Marangoni mechanism being dominant.

\subsection{Cooling experiments}
\label{cooling}
\begin{figure}[t]
\begin{center}
\includegraphics[width = 3.5in]{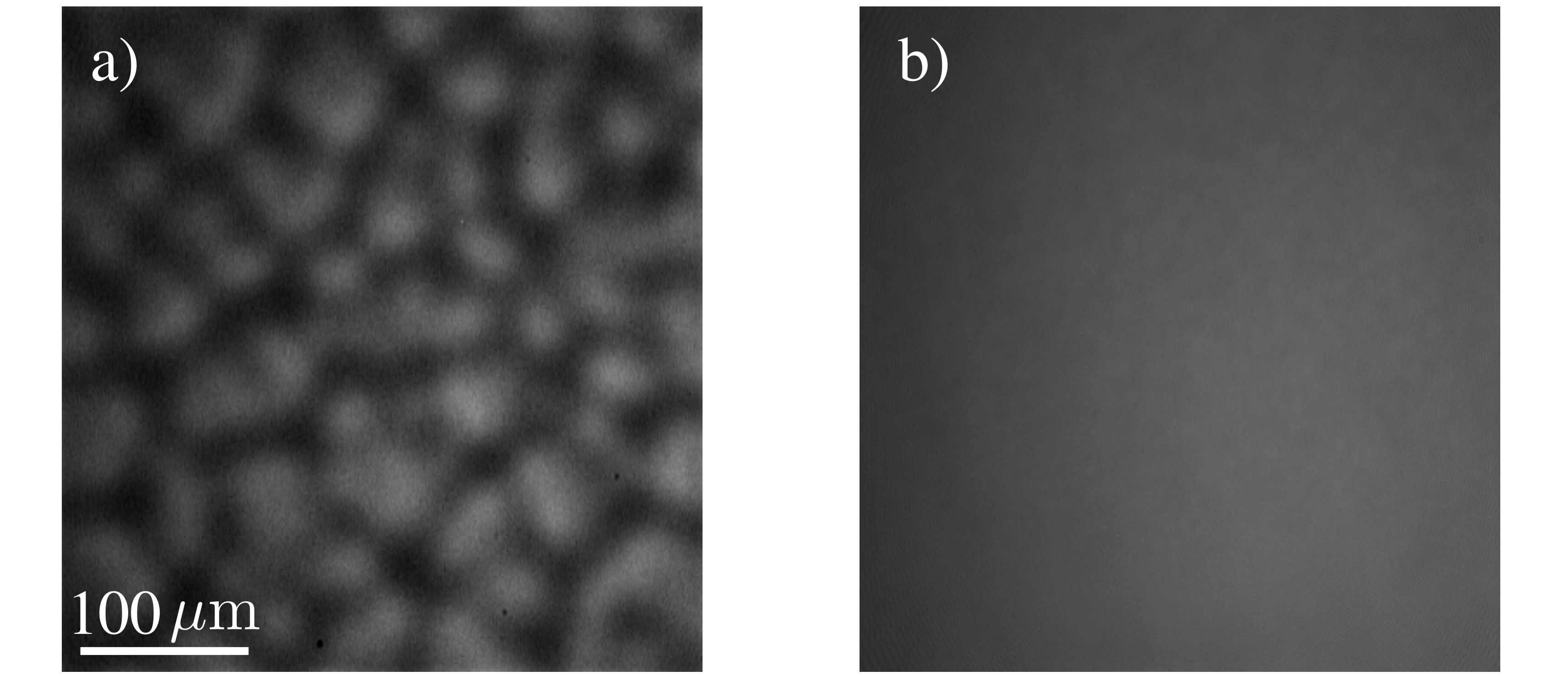}
\caption{Optical microscopy images of films spin-cast from solution of PS-PMMA with $\phi = 3.5\,\%$. (a) Films spin-cast at room temperature exhibit a flowering morphology. (b) Spincoating at  $T = 15^{\circ} $C results in uniform films. The remaining gradient in the intensity of the image is due to uneven illumination of the sample.}
\label{fig6}
\end{center}
\end{figure}
In the results presented above, we have examined how increasing the spincoating temperature affects the wavelength and amplitude of the flowering morphology. We now present the results of experiments performed with spincoating temperatures below room temperature. In fig.~\ref{fig6} are shown optical microscopy images of two films spin-cast from the same $\phi = 3.5\,\%$ solution of PS-PMMA, one at room temperature, the other at $T = 15^{\circ} $C. For the film spin-cast at room temperature, $T_{\mathrm{RT}} \approx 22^{\circ} $C, we observe a flowering morphology (fig. \ref{fig6}(a)). The film spin-cast at  $T = 15^{\circ} $C is uniform (fig. \ref{fig6}(b)). This experiment was repeated with varying solution concentrations of PS-PMMA and in each case it was found that films spin-cast at room temperature had variations in the thickness, while spincoating below room temperature resulted in uniform films.  This suggests that reducing the spincoating temperature is a simple technique to combat the formation of non-uniform spin-cast films. 

This phenomenon is easily explained in the context of the Marangoni effect. As discussed in section \ref{wavelength}, the formation of convection cells depends on the balance between surface tension and dissipation due to thermal diffusivity and viscosity. This balance is expressed in eq. (\ref{marangoni}) by the Marangoni number.   
Spincoating at lower temperatures reduces the temperature gradient across the solution caused by evaporative cooling.  According to eq. (\ref{marangoni}), $M \propto \Delta T$ so lowering the spincoating temperature decreases the Marangoni number. If the solution is sufficiently cooled, the Marangoni number will fall below the critical value, $M < M_c$, and the resulting film will be flat. 
Furthermore, we can define the critical temperature, $T_c$ as the onset temperature for Marangoni instabilities. That is for spincoating temperatures, $T>T_c$, films exhibit the flowering morphology but films spin-cast at $T<T_c$ are uniform. In fig. \ref{amp1}, $T_c$ is defined by the intersection of the best-fit straight line with the $T$ axis. For the case of PS(183k) shown in fig. \ref{amp1}, $T_c\approx 34^{\circ}$ is above room temperature  and uniform films can be spin-cast under ambient conditions. We find that $T_c$ is independent of the molecular weight over the range studied (see Electronic Supplementary Material). However as shown in fig. \ref{fig6}(a) PS-PMMA films spin-cast at room temperature are non-uniform meaning $T_c < T_{\mathrm RT}$. However, by sufficiently reducing the spincoating temperature such that $T<T_c$, we are able to prepare uniform films, as demonstrated in fig. \ref{fig6}(b).

\section{Conclusion}
\label{conclusion}
In this work we have presented a systematic study on the effect of temperature on the formation of non-uniformities in spin-cast polymer films. The `flowering morphology' observed in our experiments consists of an isotropic distribution of cells near the centre of the film, which turn to radial striations in the thickness toward  the edge of the sample. We have measured the wavelength and amplitude of the central region of the pattern as a function of the spincoating temperature. We find a linear relationship between the wavelength and average film thickness which suggests the formation of the flowering morphology is driven by a Marangoni process. 

Our results also show that the amplitude of the pattern increases linearly with spincoating temperature. We are able to explain this trend using the fact that evaporation rate increases with spincoating temperature. At higher temperatures the film has less time to flatten before vitrifying, which results in rougher films. This idea is consistent with previous studies showing that films spin-cast from more volatile solvents have rougher surfaces. Taken together, the experimental data for the amplitude and wavelength are well described by a Marangoni mechanism and not consistent with the crust formation mechanism discussed by de Gennes~\cite{deGennes:2002jl}. This is not to suggest that the crust formation mechanism does not play a role, merely that for the work discussed here the Marangoni mechanism is dominant. Further experiments should consider both mechanisms as the source of such surface structure.

Finally, we have presented a simple experimental technique to combat the formation of non-uniform films. By spincoating films at  lower temperatures we are able to entirely avoid the formation of non-uniformities in the film thickness. This result is easily explained using the Marangoni effect. The Marangoni number is proportional to the temperature gradient across the solution, $M \propto \Delta T$. Spincoating at low temperatures decreases the temperature gradient across the solution, therefore decreasing the Marangoni number. If the sample is sufficiently cooled, the Marangoni number will be lower than the critical value required to form an instability  ($M < M_c$), and the resulting film will be uniform. We emphasize that spincoating at cooler temperatures is easy to implement, though care must be take to avoid condensation of water from the air if spincoating in air. For our experiments spincoating at low temperatures was performed with a simple home-built setup; however, for an athermal solvent the same effect can be achieved by placing the polymer solution in a cooled environment several minutes prior to spinning. We are hopeful that this protocol may be useful in both research and industrial settings where the formation of non-uniform spin-cast films may be undesirable.

Financial Support for this work was provided by National Sciences
and Engineering Research Council (NSERC). JDM was supported by LabEX ENS- ICFP: ANR-10-LABX-0010/ANR-10-IDEX-0001-02 PSL.

\pagebreak
\widetext

\renewcommand{\thefigure}{S\arabic{figure}}
\setcounter{figure}{0}
\renewcommand{\theequation}{S\arabic{equation}}
\setcounter{equation}{0}

\section{Electronic Supplementary Material for ``Controlling Marangoni induced instabilities in spin-cast polymer films: how to prepare uniform films''} 

\bigskip

\section{Additional Experimental Data}
In the main text, we presented data primarily obtained from experiments performed with a $\phi$ = 2.5\% solution of PS(183k). Here we describe the results of additional measurements which probe the effects of solution concentration and molecular weight on the flowering pattern morphology. In Table \ref{tab1} is shown a complete list of the polymers used for these experiments.

\begin{table}[h]
\begin{center}
\caption{Polymers used and their molecular properties}
\label{tab1}       
\begin{tabular}{lll}
\hline\noalign{\smallskip}
polymer & $M_w$ (kg/mol) & polydispersity  \\
\noalign{\smallskip}\hline\noalign{\smallskip}
PS(8.8k) & 8.8 & 1.10 \\
PS(55.5k) & 55.5 & 1.07 \\
PS(96.5k) & 96.5 & 1.04\\
PS(135.8k) & 135.8 & 1.05 \\
PS(183k) & 183 & 1.06 \\
PS(286k) & 286 & 1.06 \\
PS(451k) & 451 & 1.10 \\
PS(490k) & 490 & 1.05 \\
PS(758.9k) & 758.9 & 1.03 \\
PS-PMMA & PS(105)-PMMA(106) & 1.13 \\
\noalign{\smallskip}\hline
\end{tabular}
\end{center}
\end{table}

\begin{figure}[t]
\begin{center}
\includegraphics[width=3.9in]{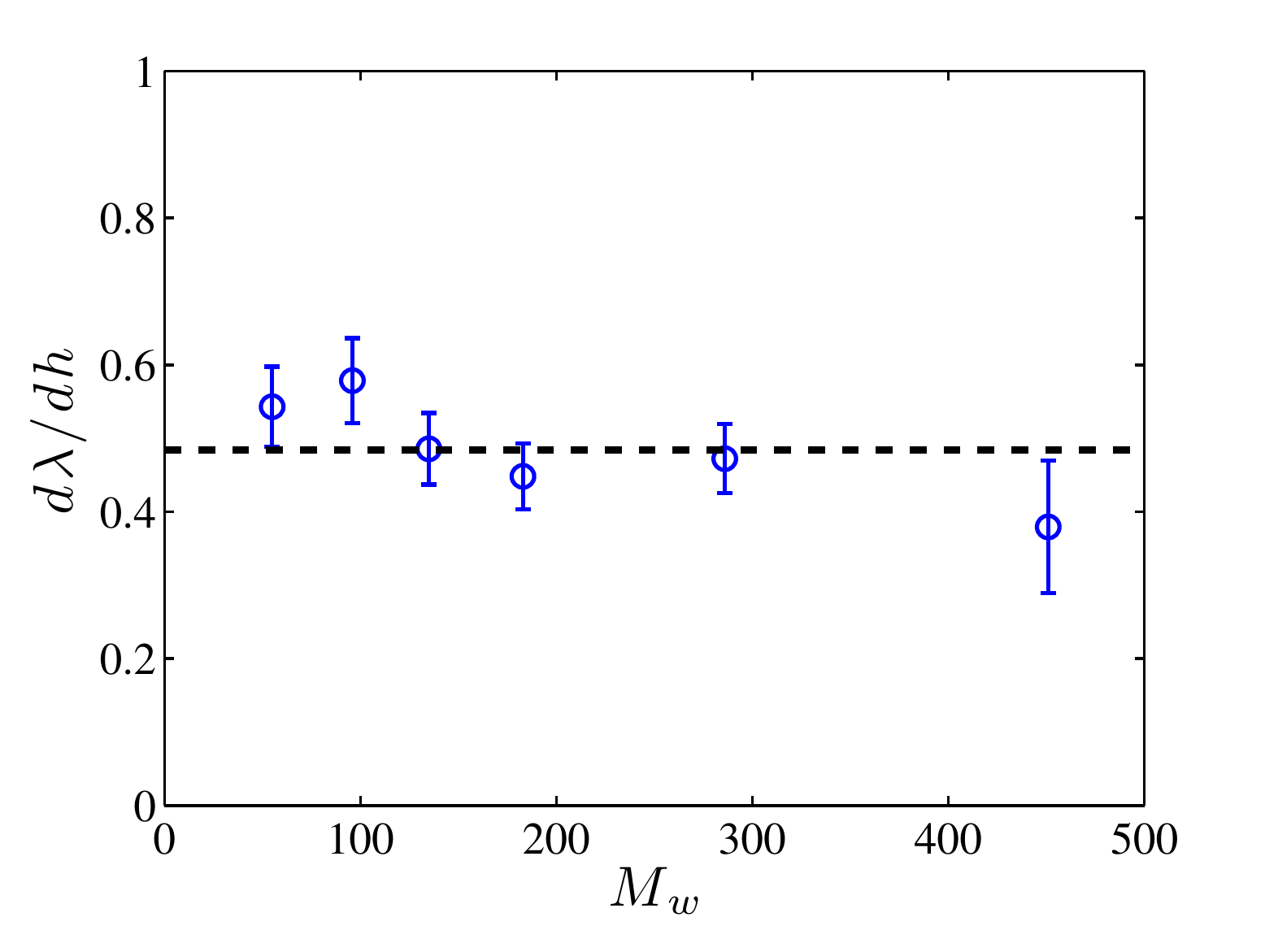}
\caption{ Measurements of $d\lambda/dh$ as a function of molecular weight. For each molecular weight, $d\lambda/dh$ is obtained by plotting the wavelength as a function of average film thickness and fitting the data to a straight line through the origin ($\lambda = mh$) as shown in Fig. 4(c) of the main text. We find $m\approx0.5$ for all molecular weights at a concentration of 2.5\,\% PS in toluene. }
\label{S1}
\end{center}
\end{figure}

First we will present data from experiments where the solution concentration was at fixed at $\phi = 2.5\%$ and the molecular weight was varied. For each molecular weight films were spincast using the same experimental procedures outlined in the main text at temperatures ranging from $15 \leq T \leq 60 \, ^{\circ} \text{C}$. For each molecular weight the wavelength was characterized as shown in Fig. 1 of the main text: we obtain both the wavelength and average film thickness as a function of spincoating temperature which together give the wavelength as a function of average film thickness. For each data set the wavelength is plotted as a function of average film thickness and the data is fit to a straight line through the origin, as exemplified in Fig. 4(c) of the main text. In Fig. \ref{S1} is shown the slope of the best fit line ($d\lambda/dh$) as a function of molecular weight. For all molecular weights we find that $\lambda \approx 0.5h$.

For the experiments described above we also measured the amplitude as a function of spincoating temperature for each molecular weight. From these data sets we are able to extract the critical temperature, $T_c$, as a function of molecular weight (at a constant solution concentration $\phi = 2.5\%$) using the method shown in Fig. 5 of the main text. The amplitude is plotted as a function of spincoating temperature and the data is fit to a straight line. The extrapolation of the best fit line through the temperature axis gives $T_c$. Fig. S2 shows the critical temperature as a function of molecular weight. We observe that within experimental error, $T_c$  is independent of molecular weight. For solutions of polystyrene with $\phi = 2.5\,\%$ the critical temperature is found to be $T_c \approx 31 \pm 5^{\circ} \text{C }$. 

\begin{figure}[b]
\begin{center}
\includegraphics[width=3.9in]{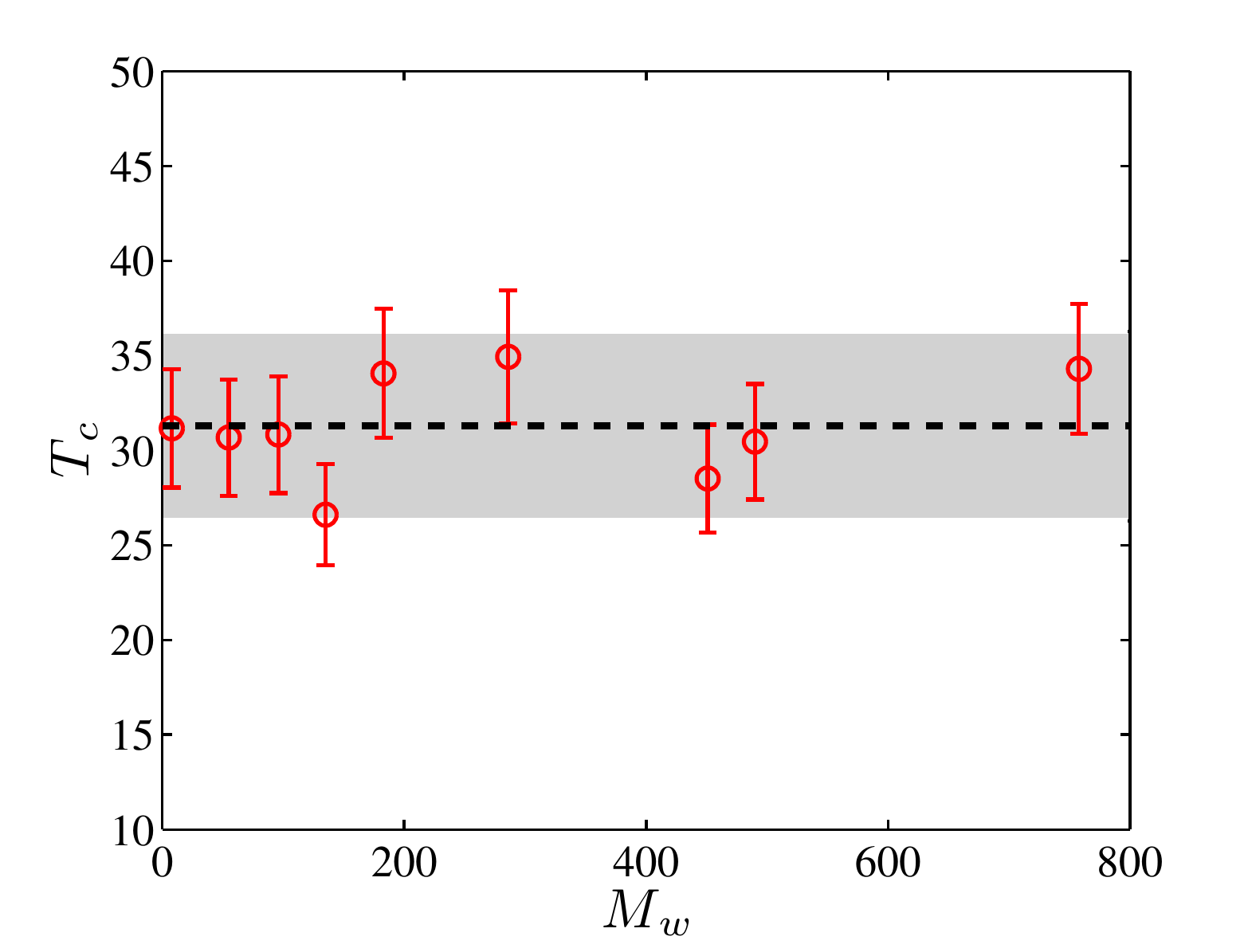}
\caption{ $T_c$ as a function of molecular weight. Each data point was obtained as follows. The amplitude of the flowering pattern as a function of spincoating temperature was measured using imaging ellipsometry for temperatures ranging from $15 \leq T \leq 60 \, ^{\circ} \text{C}$ and the data was plotted as exemplified in Fig. 5 of the main text. From such plots, the extrapolation of the best fit line through the temperature axis gives $T_c$. $T_c \approx 31 \pm 5 ^{\circ} \text{C}$ for all of the molecular weights presented here at a concentration of 2.5\,\% PS in toluene. }
\label{S2}
\end{center}
\end{figure}

The observation that $T_c$ is larger than room temperature is consistent with what is typically observed in experimental laboratories. Spincoating polystyrene from toluene at room temperature results in flat films. The scatter in the data can be attributed to variations in ambient temperature and humidity. Experiments were performed on different days and slight variations in ambient conditions can effect properties such as the solvent evaporation rate. Despite our best efforts we were unable to control for variations in ambient conditions. 

\begin{figure}[t]
\begin{center}
\includegraphics[width=3.9in]{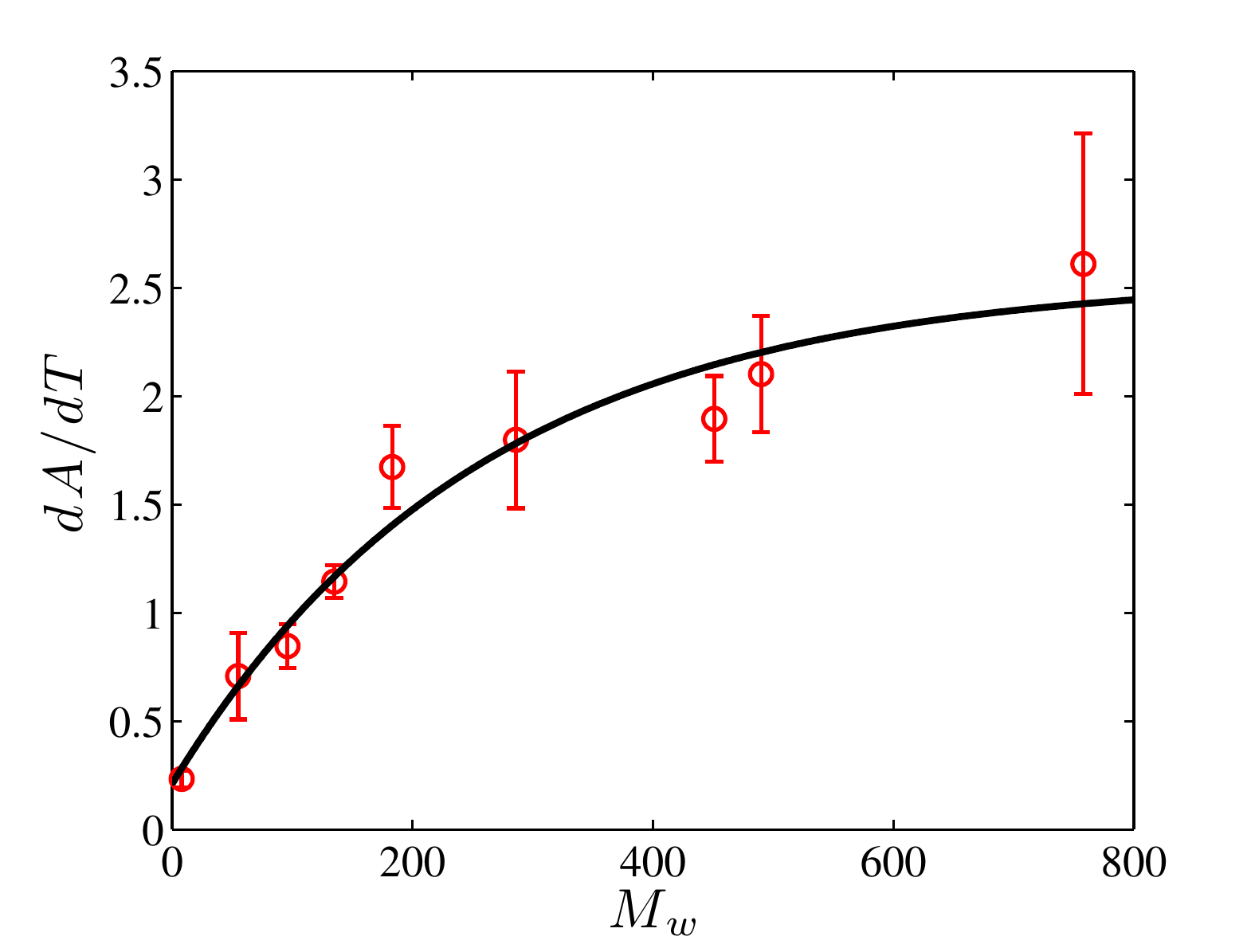}
\caption{ $dA/dT$ as a function of molecular weight. The black line as an exponential and only intended as a guide to the eye.}
\label{S3}
\end{center}
\end{figure}

From plots of the amplitude as a function of spincoating temperature (eg. Fig. 5 of the main text) we also extract the slope of the best fit line.  In Fig.~\ref{S3} is shown the slope of the best fit line ($dA/dT$) as a function of molecular weight. 

\begin{figure}[b]
\begin{center}
\includegraphics[width=3.9in]{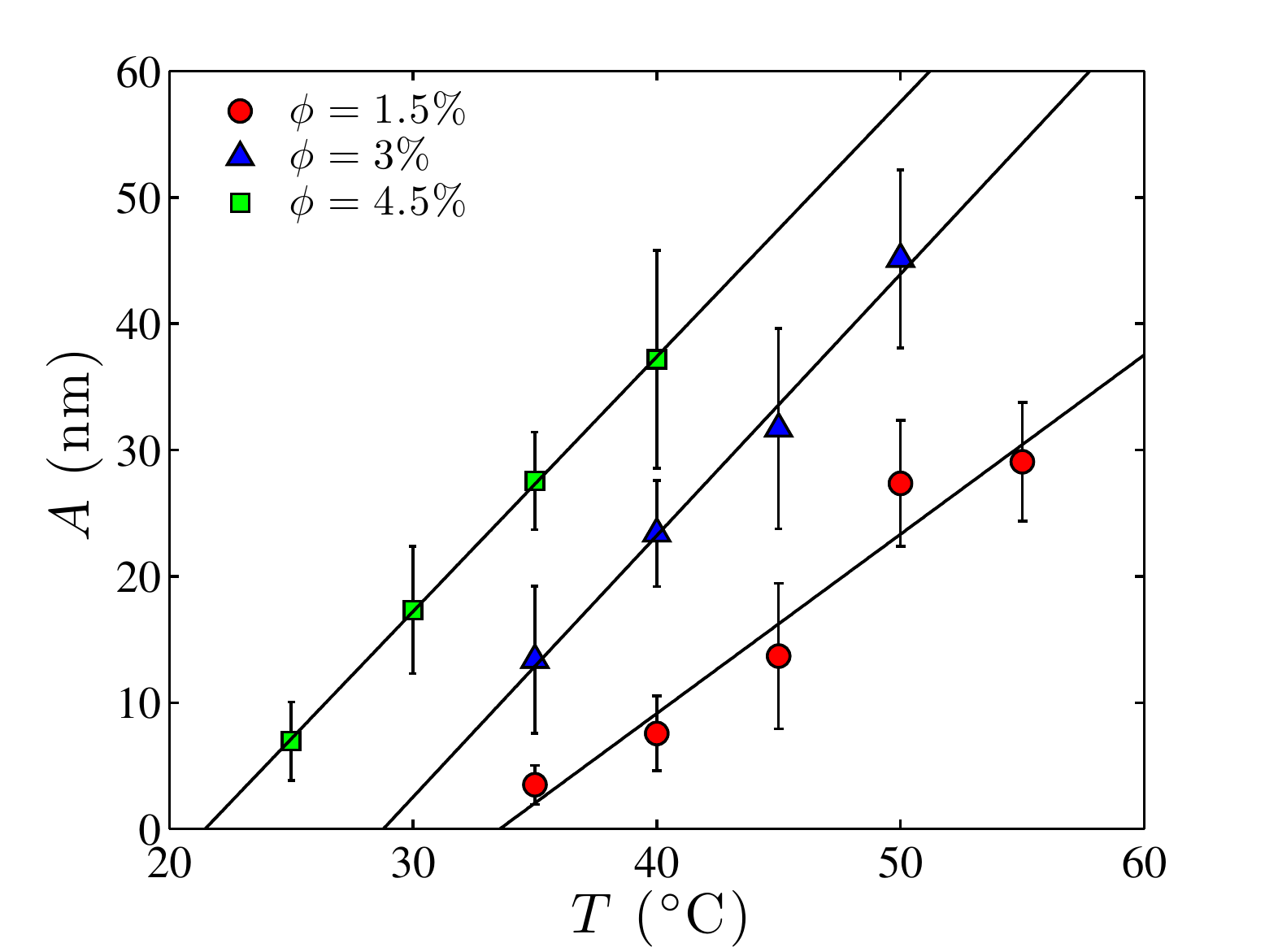}
\caption{ Measurements of amplitude as a function of spincoating temperature for varying solution concentrations of PS(183k). $T_c$ decreases with increasing solution concentration.}
\label{S4}
\end{center}
\end{figure}

Finally we present the results of experiments probing the effect of solution concentration on the flowering morphology. In Fig.~\ref{S4} is plotted the amplitude as a function of temperature for films spincast from solutions of PS(183k) with $1.5 \leq \phi \leq 3\%$. We observe a reduction in $T_c$ with increasing solution concentration. This data is consistent with the common empirical observation that as film thickness increases it becomes more difficult to spin coat uniform films~\cite{Spangler:1990ux,Strawhecker:2001ch} under ambient conditions. We note that increasing the solution concentration results in thicker films and we therefore expect a lower $T_c$ for a more concentrated solution.

\end{document}